# Electric Potential Due to a System of Conducting Spheres

Philip T. Metzger
NASA Granular Mechanics and Regolith Operations Laboratory
Mail Code: KT-D3
Kennedy Space Center, FL 32899
Philip.T.Metzger@nasa.gov

John E. Lane
ASRC Aerospace, MS: ASRC-24
Kennedy Space Center, FL 32899
John.E.Lane@ksc.nasa.gov

**Abstract:**  Equations describing the complete series of image charges for a system of conducting spheres are presented.  The method of image charges, originally described by J. C. Maxwell in 1873, has been and continues to be a useful method for solving many three dimensional electrostatic problems.  Here we demonstrate that as expected when the series is truncated to any finite order *N*, the electric field resulting from the truncated series becomes qualitatively more similar to the correct field as *N* increases.  A method of charge normalization is developed which provides significant improvement for truncated low order solutions. The formulation of the normalization technique and its solution via a matrix inversion has similarities to the method of moments, which is a numerical solution of Poisson's equation, using an integral equation for the unknown charge density with a known boundary potential.  The last section of this paper presents a gradient search method to optimize a set of *L* point charges for *M* spheres.  This method may use the image charge series to initialize the gradient search.  We demonstrate quantitatively how the metric can be optimized by adjusting the locations and amounts of charge for the set of points, and that an optimized set of charges generally performs better than truncated normalized image charges, at the expense of gradient search iteration time.

# INTRODUCTION

The first year graduate physics student is likely to be introduced to the problem of calculating the electric field surrounding a conducting sphere in the presence of a point charge during the first few weeks of a standard course in electromagnetism. When this problem is extended to include a cluster of spheres, things get interesting, as well as more difficult. One application of this problem may be found in nuclear physics. Nuclear physicists sometimes need to sum the probability of nuclei breaking into every possible configuration of clusters, each cluster being modeled by a set of charged spheres, and this entails a large number of configurations, each of which must be solved individually. The probability of the nuclei breaking requires a calculation of the stored energy of the electric charges, which depends upon their actual distribution on the spheres, and it is computationally expensive for such a large number of configurations. Therefore, they use the image charge method truncating the series of image charges for computational efficiency, but at the cost of some accuracy.

Another application involves a proposed spacecraft [electrostatic radiation shield](#), made up of a cluster of conducting spheres surrounding the spacecraft. Similarly, a [lunar radiation shield](#) study incorporated conducting spheres of various sizes and potentials. In order to simulate the benefits of a radiation shield configuration, the electric field is needed at every location around the spacecraft in order to calculate the trajectories of the charged particles that constitute the cosmic radiation in space. The electric field throughout space depends upon the actual distribution of charge on the spheres, and it is computationally expensive to (first) solve for the actual distribution of charge, and (second), integrate the contributions to the electric field in each location of space resulting from all the portions of the surfaces of the charged spheres. For computational efficiency, we use the image charge method truncating the series of image charges so that only a finite set of point charges contribute to the electric field in all locations of space around the spheres, and thus summing these contributions is a simple sum over only a finite set of point charges rather than an integral over a set of surfaces.

In many applications it is computationally expensive to solve the exact distribution of charges since the charge distribution on the spheres is not uniform when multiple charged spheres interact with one another. As spheres move increasingly close to one another, the charges on each sphere are pushed around by the electric fields of adjacent spheres. Since the electric fields from adjacent spheres are also changing, as their own charge distributions are perturbed, the final distribution of charge on the spheres becomes difficult to calculate.

A mathematical technique that can lead to an exact solution in many electrostatic problems is based on conformal mapping in the complex plane. Solving Laplace's equation by conformal mapping has been primarily restricted (until recently) to two-dimensional problems, or to three-dimensional problems that have rotational symmetry, or to cases where the Separation of Variables method can be applied [5, 6]. These cases do not encompass the problem solved in this paper, charged spheres placed arbitrarily in three-dimensional space, which is not generally reducible to a two-dimensional problem or amenable to the Separation of Variables. However, the techniques of conformal mapping are advancing at a rapid pace, extending the scope of problems that can be solved by this method [7-9]. At least some of these advances involve infinite products and/or infinite series as a part of the solution, and therefore may provide no advantage in computing a numerical solution in electrostatics. However, that waits to be seen, and this paper makes no attempt to evaluate the rapidly advancing methods in conformal mapping.

Another technique that is of general usefulness in three-dimensional electrostatic problems is the Method of Moments [10, 11]. The Method of Moments can be used to solve Poisson's equation by finding the unknown charge density on the surface of a conductor when the potential of the conductor is known. Fairly large matrices can result which are inverted to find the surface charge density. A key strategy to this method is find useful basis functions that can be solved analytically, thus reducing the number of matrix elements that need to be solved numerically. The Method of Moments is a powerful method which can be used to solve a large variety of electrostatic, as well as general electromagnetic problems. For the specific problem discussed in this paper, the Method of Images leads to a more direct approach, which is simpler conceptually and requires less complex computer code to implement as compared to the Method of Moments. Also, since the problem addressed by this paper generally involves three-dimensional geometry without rotational symmetry, the Method of Images is a simpler solution method compared to a method based on conformal mapping into the complex plane.

**IMAGE CHARGE SOLUTIONS**

The Method of Images [1, 2] is a convenient procedure for finding the electric potential due to a system of conductors and point charges without having to solve a differential equation, where the solution is guaranteed to be a solution of Laplace's equation in the *exterior region*. The work involved with this method is to simply match boundary conditions via vector algebra using a finite or infinite series of point charge solutions. However, finding the necessary image point charges is not necessarily trivial. Since image charge solutions generally involve an infinite series of images, except in very simple cases, truncation of the series will always result in a less than perfect solution.

**Point Charge Near a Conducting Sphere**

The simplest example problem that will help lead into the general problem of calculating the electric potential due to a system of conducting spheres, is that of a single point charge outside of a conducting sphere of radius $a$ held a constant potential $V_0$. This problem is presented in detail in Jackson [2], but will be repeated below in brief to serve as an introduction into the more difficult cases to be considered next.

The point charge has a charge $q$ and is a distance $d$ from the sphere's center, noting that $d > a$. The potential can be written as that due to one real charge plus two image charges:

$$U(\mathbf{r}) = \frac{1}{4\pi\varepsilon_0}\left(\frac{q_0}{r} + \frac{q_1}{|\mathbf{r}-b\mathbf{e}_x|} + \frac{q}{|\mathbf{r}-d\mathbf{e}_x|}\right)$$
$$= \frac{aV_0}{r} + \frac{q}{4\pi\varepsilon_0}\left(\frac{\gamma}{|\mathbf{r}-b\mathbf{e}_x|} + \frac{1}{|\mathbf{r}-d\mathbf{e}_x|}\right) \quad (1)$$

where the charges are placed on the *x*-axis for convenience, without loss of generality and where $\mathbf{e}_x$ denotes the unit vector along the *x*-axis, as shown in Fig. (**1**). The first term is that due to an image charge at the center of the sphere, proportional to sphere's potential. The second term is a single image charge that is induced at $\mathbf{r} = b\mathbf{e}_x$ by the proximity of the real charge (the third term), located at $\mathbf{r} = d\mathbf{e}_x$. Note that $b < a$. The magnitude of the image charge is designated as $\gamma q$.

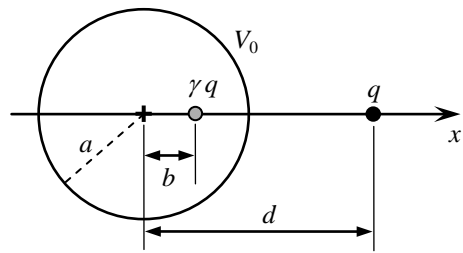

**Fig. (1).** Point charge $q$ a distance $d$ from a conducting sphere of radius $a$.

The solution for the position and magnitude of the image charge, as described in the references [2-4], is:

$$b = a^2/d \quad (2a)$$
$$\gamma = -a/d \quad (2b)$$

A demonstration of the single conducting sphere and single point charge is shown in Fig. (**2a**) with $a$ = 1.2, $d$ = 2.5, and $V_0$ = 0, with the image charge parameters arbitrarily set to $\gamma$ = -1 and $b$ = 0. In Fig. (**2b**), $\gamma$ and $b$ are set to the correct values, according to Eq. (2).

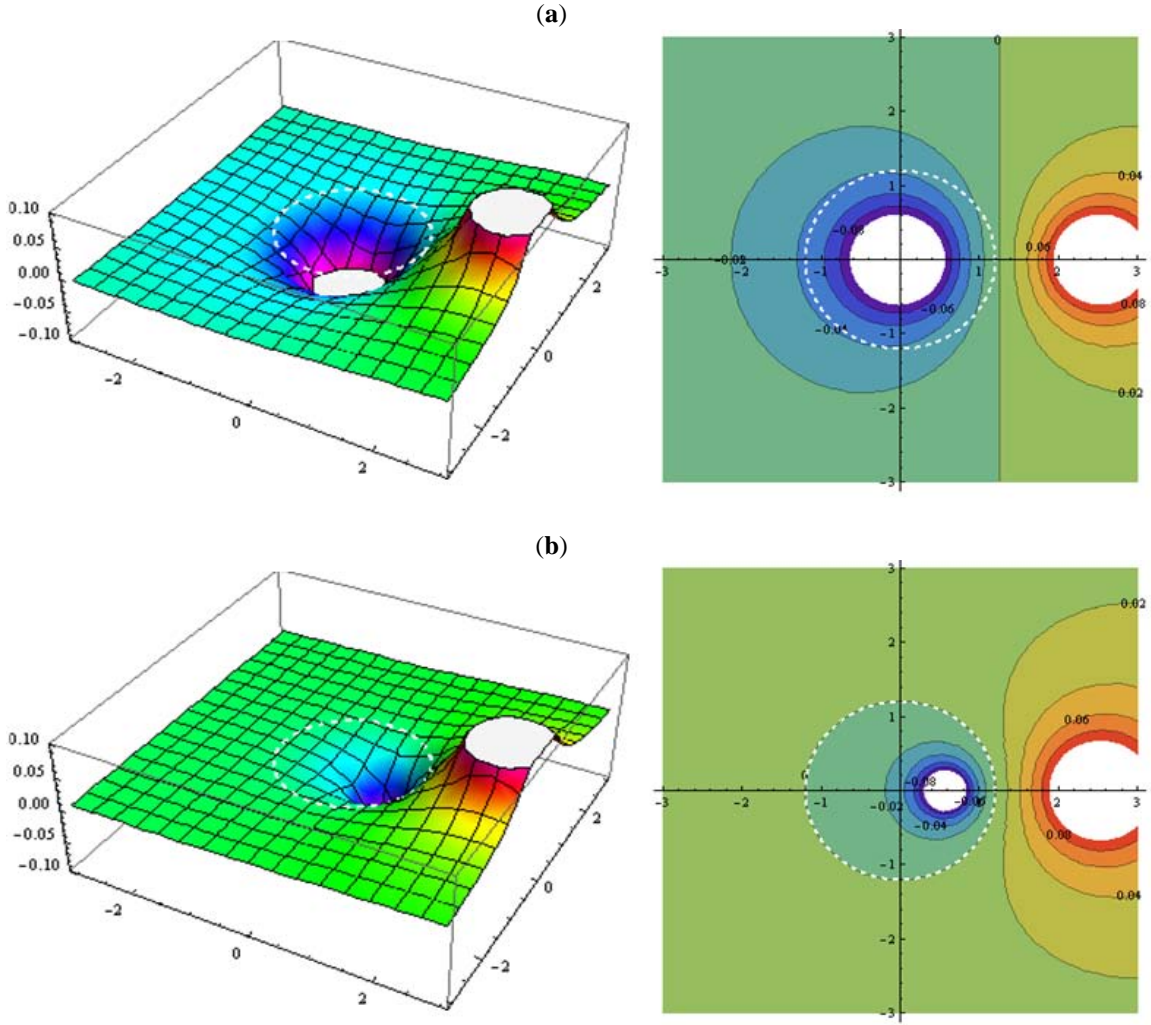

**Fig. (2).** A single point charge near a conducting sphere at $z$ = 0 along the $x$-$y$ plane, using Eq. (1) with $V_0$ = 0. The potential U(**r**) is pseudo colorized with a 3D plot on the left and corresponding contour plot on the right. The dotted white circle represents the correct potential on the surface of the sphere. (**a**) $\gamma$ = -1, $b$ = 0; (**b**) $\gamma$ and $b$ are set according to Eq. (2).

**Image Charges for Two Conducting Spheres**

Fig. (**3**) shows two conducting spheres at different constant potentials, $V_1$ and $V_2$. The electric field potential can again be solved by use of image charges (see reference [12] for historical insights). The result is an infinite series of image charges inside of each sphere where the magnitude decreases with increasing order. As the order increases, positions of the image charges move closer to the inside surface of the sphere. No image charges appear outside of the spheres, in accordance with Laplace's equation and the uniqueness theorem.

The strategy (as well as similar notation) behind solving this problem is based on Jackson [2]. Because the electric field potential can be decomposed into a sum of fields due to image point charges, the solution of the single sphere with a single external point charge, as described in the previous section, can be applied in an iterative fashion to solve this problem.

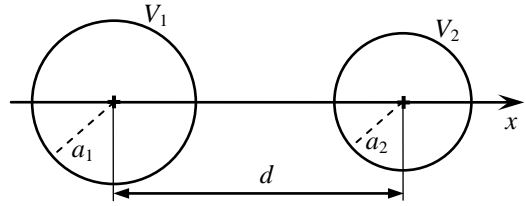

**Fig. (3).** Two conducting spheres at different constant potentials.

The zeroth order image charges are located at the center of each sphere:

$$q_1(0) = 4\pi\varepsilon_0 a_1 V_1 \qquad (i = 1) \tag{3a}$$

$$q_2(0) = 4\pi\varepsilon_0 a_2 V_2 \qquad (i = 2) \tag{3b}$$

As the distance d between spheres increases, the effect of the higher order terms is diminished. For many applications where the sphere separation is much greater than the sphere radius, the zeroth order solution will suffice and higher order terms will not be necessary.

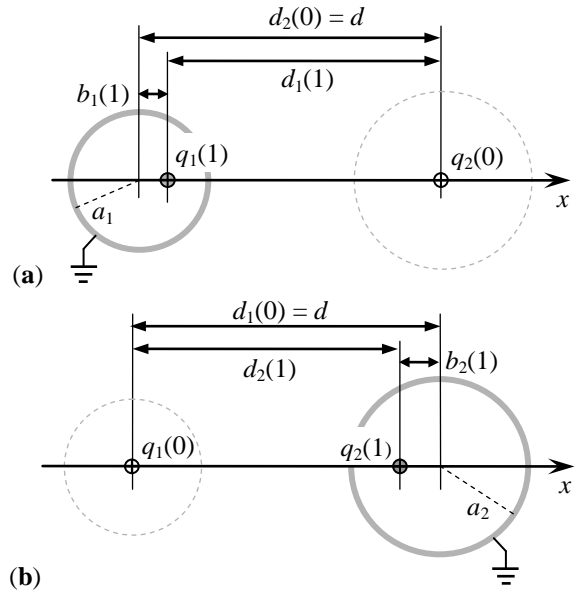

**Fig. (4).** First order image charges: (**a**) $i = 1$; (**b**) $i = 2$.

First order image charges are determined one at a time for each $i$th sphere ($i = 1, 2$ in the current case). The diagram for determining the $i = 1$ first order image charge is shown in Fig. (**4a**). Comparing this figure with Fig. (**1**) with the sphere grounded and Eqs. (2), lead immediately to the value and location of the first order charge:

$$q_1(1) = -\frac{a_1}{d_2(0)} q_2(0)$$

$$b_1(1) = \frac{a_1^2}{d_2(0)} \qquad (i = 1) \tag{4a}$$

$$d_1(1) = d - b_1(1)$$

Similarly, the $i = 2$ first order image is determined from Fig. (**4b**):

$$q_2(1) = -\frac{a_2}{d_1(0)} q_1(0)$$

$$b_2(1) = \frac{a_2^2}{d_1(0)} \qquad (i = 2) \tag{4b}$$

$$d_2(1) = d - b_2(1)$$

This process continues in an iterative fashion, as demonstrated in Figs. (**5a-5i**), which shows the step by step determination of zeroth order through third order terms. Note that grounding a sphere is perfectly valid, since the zeroth order image charges account for the sphere's potential.

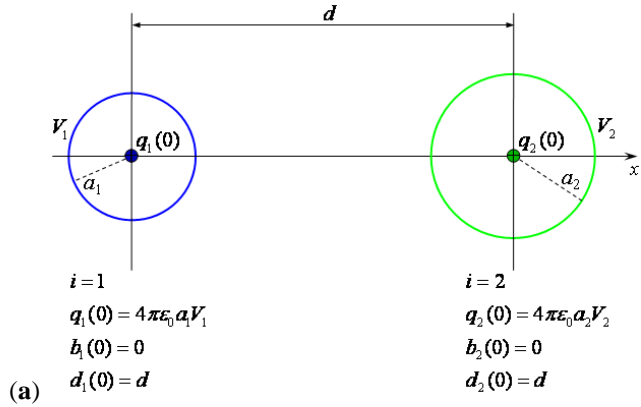

(a)

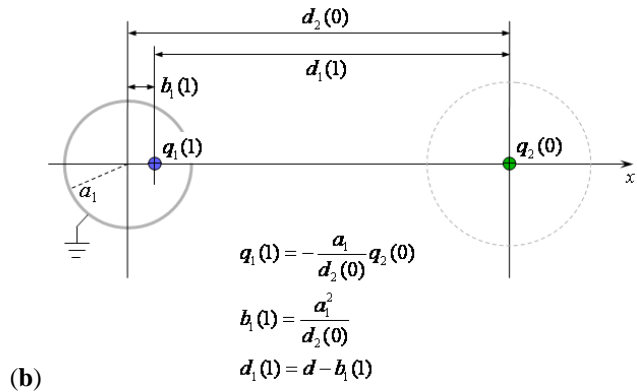

(b)

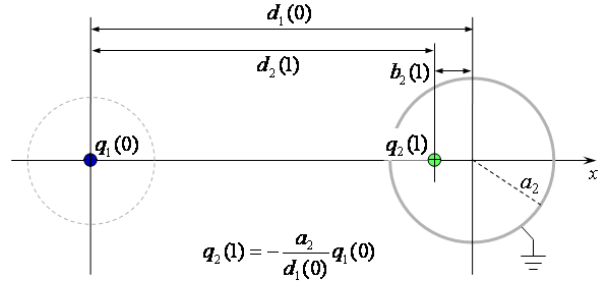

(**c**)
$$q_2(1) = -\frac{a_2}{d_1(0)} q_1(0)$$
$$b_2(1) = \frac{a_2^2}{d_1(0)}$$
$$d_2(1) = d - b_2(1)$$

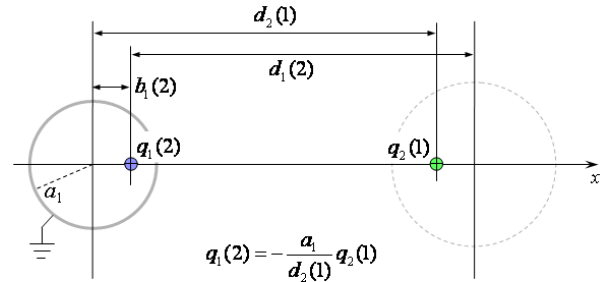

(**d**)
$$q_1(2) = -\frac{a_1}{d_2(1)} q_2(1)$$
$$b_1(2) = \frac{a_1^2}{d_2(1)}$$
$$d_1(2) = d - b_1(2)$$

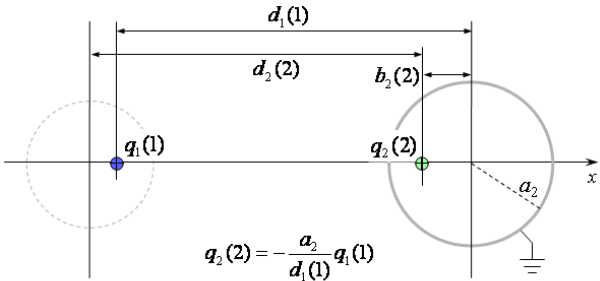

(**e**)
$$q_2(2) = -\frac{a_2}{d_1(1)} q_1(1)$$
$$b_2(2) = \frac{a_2^2}{d_1(1)}$$
$$d_2(2) = d - b_2(2)$$

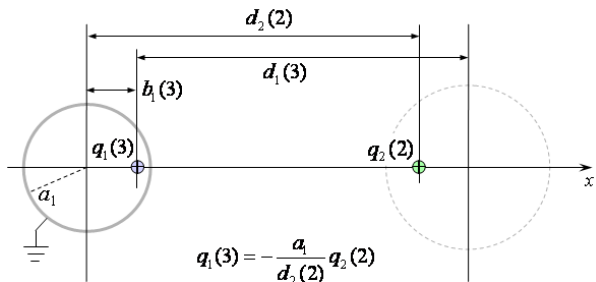

(**f**)
$$q_1(3) = -\frac{a_1}{d_2(2)} q_2(2)$$
$$b_1(3) = \frac{a_1^2}{d_2(2)}$$
$$d_1(3) = d - b_1(3)$$

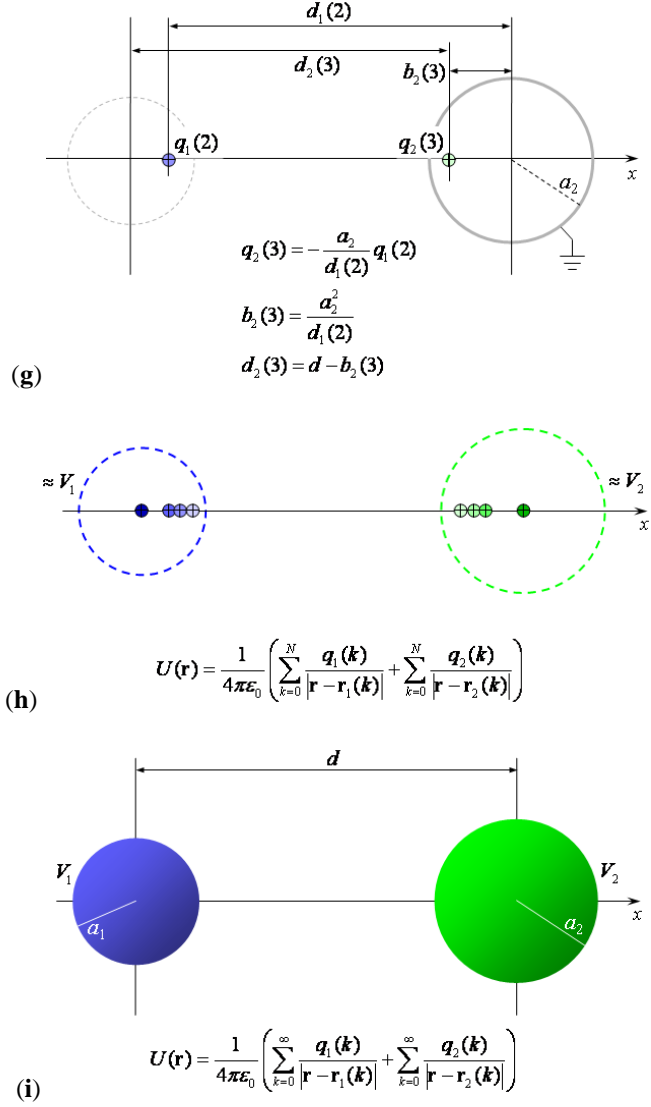

**Fig. (5).** Demonstration of image charge series for two conducting spheres using Eq. (5): (**a**) zeroth order term; (**b**) first order term, $i = 1$; (**c**) first order term, $i = 2$; (**d**) second order term, $i = 1$; (**e**) second order term, $i = 2$; (**f**) third order term, $i = 1$; (**g**) third order term, $i = 2$; (**h**) sum of all $N$th order terms from Eq. (7); (**i**) exact solution using infinite series.

The magnitude and positions of the $k$th image charges belonging to the $i$th sphere can be summarized by the following set of iterative formulas:

$$q_i(k) = -\frac{a_i}{d_j(k-1)} q_j(k-1)$$

$$b_i(k) = \frac{a_i^2}{d_j(k-1)} \quad (5)$$

$$d_i(k) = d - b_i(k)$$

where $i = 1, 2$ and $j = 1, 2$, with $j \neq i$. As is evident from Figs. (**4**) and Figs. (**5**), $b_i(k)$ is the offset of the $k$th order image charge from the $i$th sphere's center in the direction of the $j$th sphere. The parameter $d_i(k)$ is redundant, even though it is useful in developing the recursive set of formulas summarized in Eq. (5). Eliminating $d_i(k)$ results in a minimum two parameter set, describing the magnitude and position of all image charges in the two-sphere system:

$$q_i(k) = -\frac{a_i}{d - b_j(k-1)} q_j(k-1)$$

$$b_i(k) = \frac{a_i^2}{d - b_j(k-1)}$$

(6)

where $b_i(0) = 0$. The electric potential for the two sphere system, using terms up to $N$th order is now:

$$U(\mathbf{r}) = \frac{1}{4\pi\varepsilon_0} \left( \sum_{k=0}^{N} \frac{q_1(k)}{|\mathbf{r} - \mathbf{r}_1(k)|} + \sum_{k=0}^{N} \frac{q_2(k)}{|\mathbf{r} - \mathbf{r}_2(k)|} \right)$$

(7)

where the zeroth order charges are given by Eqs. (**3**). In the special case where sphere-1 is centered at the origin of the $xyz$ coordinate system and sphere-2 is centered at a distance $d$ along the $x$-axis, the location vectors in Equation (7) simply to:

$$\mathbf{r}_1(k) = b_1(k)\mathbf{e}_x, \qquad \mathbf{r}_2(k) = d_2(k)\mathbf{e}_x$$

(8)

where $\mathbf{e}_x$ is the unit vector along the $x$-axis.

**Charge Normalization of Two Conducting Spheres**

The previous section shows that by increasing the order of the solution $N$, the voltage $V_k$ on the surface of the $k$th sphere converges to the correct constant value at all points on the surface. For any specific practical application, one simply decides on a value $N$ that should satisfy the precision requirements for the application. For a given precision, the number of image charges needed to achieve that precision can be reduced by an order or more using an optimization method based on normalization of the charge. In this way for example, a normalized first order solution of two image charges per sphere should be approximately more accurate than a second order solution without normalization. The concept of parameter optimization is common to many fields involving use of a truncated infinite series.

Charge normalization is a special case of optimization where the voltage specified in the zeorth order charges, Eqs. (3a) and (3b) are modified by a scaling parameter:

$$q_1(0) = 4\pi\varepsilon_0 a_1 (\alpha_1 V_1)$$

(9a)

$$q_2(0) = 4\pi\varepsilon_0 a_2 (\alpha_2 V_2)$$

(9b)

By adjusting the $\alpha$'s, the truncated solution given by Eq. (7) can be optimized so that the voltage on the sphere surface $U(\mathbf{r} = \mathbf{r}_k + a_k \mathbf{e}) \to V_k$ for $k = 1$ and 2 (the next section will consider the general case of any number of spheres). This can be summarized by the following matrix equation:

$$\begin{pmatrix} U(\mathbf{r}_1 + a_1\mathbf{e}) \\ U(\mathbf{r}_2 + a_2\mathbf{e}) \end{pmatrix} = \mathbf{C} \cdot \begin{pmatrix} \alpha_1 \\ \alpha_2 \end{pmatrix}$$
$$= \begin{pmatrix} V_1 \\ V_2 \end{pmatrix}$$
(10)

where the components of **C** are obtained from Eq. (7) by re-ordering terms so that the $q_k(0)$'s can be grouped and factored. The components of **C** are then:

$$c_{11}(\theta,\phi) = \sum_{\substack{k=0 \\ k \text{ even}}}^{N} \frac{q_1(k)}{|a_1\mathbf{e} - b_1(k)\mathbf{e}_x|} + \sum_{\substack{k=1 \\ k \text{ odd}}}^{N} \frac{q_2(k)}{|a_1\mathbf{e} - d_2(k)\mathbf{e}_x|}$$
(11a)

$$c_{12}(\theta,\phi) = \sum_{\substack{k=1 \\ k \text{ odd}}}^{N} \frac{q_1(k)}{|a_1\mathbf{e} - b_1(k)\mathbf{e}_x|} + \sum_{\substack{k=0 \\ k \text{ even}}}^{N} \frac{q_2(k)}{|a_1\mathbf{e} - d_2(k)\mathbf{e}_x|}$$
(11b)

$$c_{21}(\theta,\phi) = \sum_{\substack{k=0 \\ k \text{ even}}}^{N} \frac{q_1(k)}{|a_2\mathbf{e} + d_1(k)\mathbf{e}_x|} + \sum_{\substack{k=1 \\ k \text{ odd}}}^{N} \frac{q_2(k)}{|a_2\mathbf{e} + d_1(k)\mathbf{e}_x|}$$
(11c)

$$c_{22}(\theta,\phi) = \sum_{\substack{k=1 \\ k \text{ odd}}}^{N} \frac{q_1(k)}{|a_2\mathbf{e} + d_1(k)\mathbf{e}_x|} + \sum_{\substack{k=0 \\ k \text{ even}}}^{N} \frac{q_2(k)}{|a_2\mathbf{e} + b_2(k)\mathbf{e}_x|}$$
(11d)

where the unit vector **e** is a function of the spherical angles $\theta$ and $\phi$:

$$\mathbf{e} = \begin{pmatrix} \sin\theta\cos\phi \\ \sin\theta\sin\phi \\ \cos\theta \end{pmatrix}$$
(12)

The constants computed in Equations (11a) through (11d) apply to a single arbitrary point on the surface of a sphere determined by the angles $\theta$ and $\phi$. These values need to be integrated over the entire sphere surface in order to complete the normalization procedure. A simple approximation to that integration can be obtained by averaging the $c_{ij}$'s over random values of the angles, as follows:

$$\mathbf{C} = \begin{pmatrix} C_{11} & C_{12} \\ C_{21} & C_{22} \end{pmatrix} = \begin{pmatrix} \frac{1}{N_R}\sum_{n=1}^{N_R} c_{11}(s_n\pi, u_n 2\pi) & \frac{1}{N_R}\sum_{n=1}^{N_R} c_{12}(s_n\pi, u_n 2\pi) \\ \frac{1}{N_R}\sum_{n=1}^{N_R} c_{21}(s_n\pi, u_n 2\pi) & \frac{1}{N_R}\sum_{n=1}^{N_R} c_{22}(s_n\pi, u_n 2\pi) \end{pmatrix}$$
(13)

where $u_n$ is a uniform random number in the interval of 0 to 1, and $s_n$ is a sinusoidally distributed random number in the interval of 0 to 1. A sinusoidally distributed random number $s_n$ can be generated from a normally distributed random number $u_n$ by the following algorithm:

$$s_n = \frac{\cos^{-1}(1 - 2u_n)}{\pi}$$
(14)

To solve for the $\alpha$'s, Equation (10) is inverted:

$$\begin{pmatrix} \alpha_1 \\ \alpha_2 \end{pmatrix} = \mathbf{C}^{-1} \cdot \begin{pmatrix} V_1 \\ V_2 \end{pmatrix} \tag{15}$$

An example of the benefit of optimization by charge normalization is revealed by Figs. (**6**), where $d = 3.5$, $a_1 = 1.5$, $a_2 = 1$, $V_1 = 0.6$, and $V_2 = 0.8$. Again the dotted white circles represent the location of the correct values of the sphere voltages $V_1$ and $V_2$ in the 3D plot space. As can be seen by the voltage contour plots, alignment with the dotted circles representing the spheres, an $N-1$ order solution with charge normalization optimization appears as good or better than an $N$th solution without optimization.

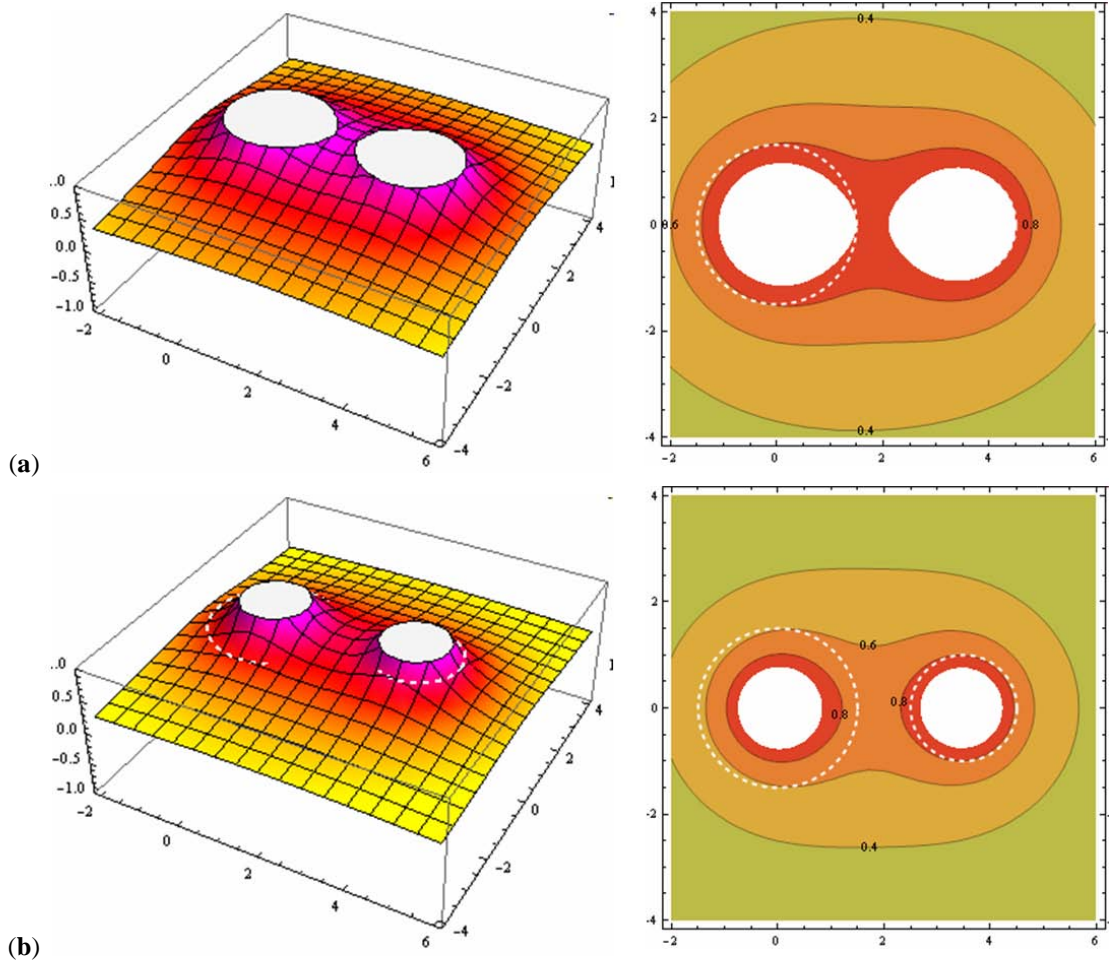

(**a**)

(**b**)

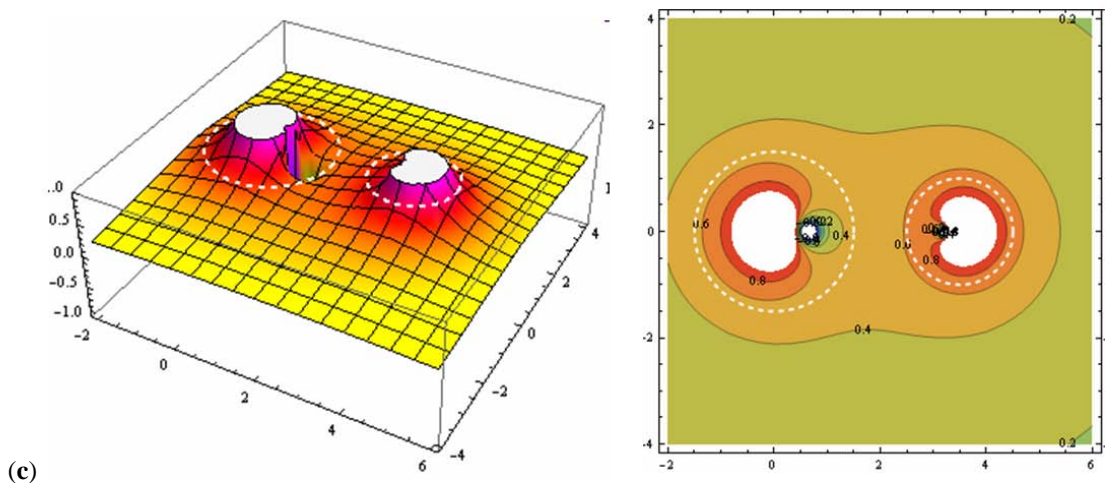

(c)

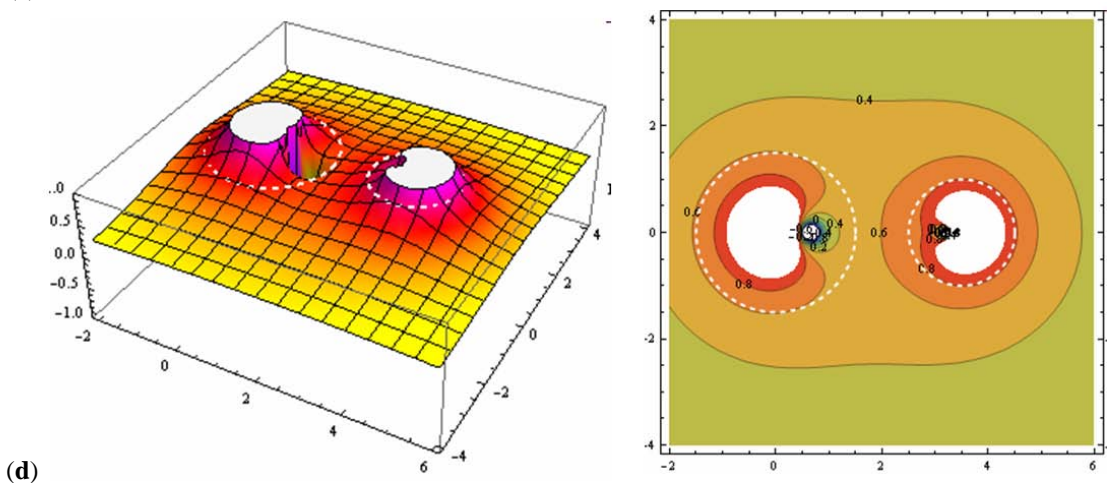

(d)

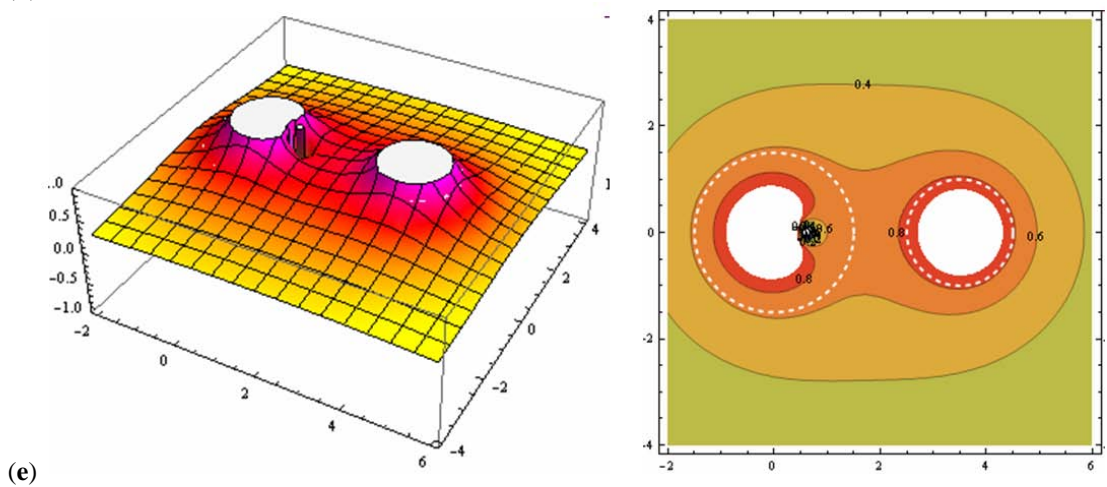

(e)

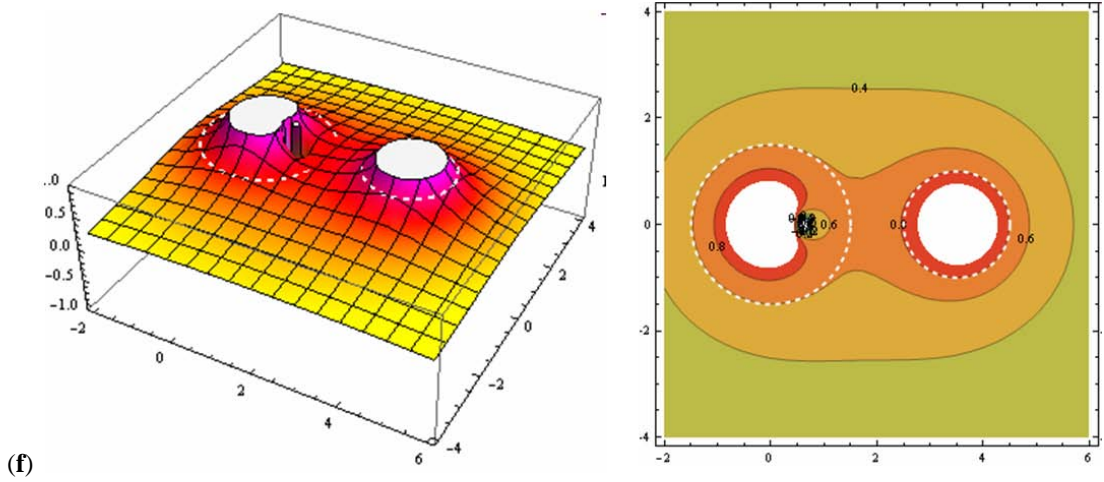

(f)

**Fig. (6).** Comparison of zeroth order through second order image method electric field potentials of two conducting spheres. The potential $U(\mathbf{r})$ is pseudo colorized, and is computed from Eq. (7) at $z = 0$ along the $x$-$y$ plane without and with charge normalization using Eqs. (9) and (15). The dotted white circle represents the correct potential on the sphere surfaces: (**a**) zeroth order image term only; (**b**) zeroth order term with normalization; (**c**) include first image order term; (**d**) including first order terms with normalization; (**e**) includes second order image order term; (**f**) including second order terms with normalization.

## IMAGE CHARGES FOR *M* CONDUCTING SPHERES

The concepts of the previous section can now be extended to the general case of *M* spheres. In order to proceed, the notation of Jackson [2] will be abandoned and replaced with a notation more suitable for the general case. Jackson's notation is efficient for representing the general case of *N*th order, but practically restricted to only two spheres. This new notation is efficient for representing *M* spheres, but tends to become unwieldy for $N > 3$. Never-the-less, this notation provides the a roadmap to those formulas describing higher order image terms.

The zeroth order image charges, analogous to the notation of Eqs. (3), can be alternately expressed by:

$$q_i = 4\pi\varepsilon_0 a_i V_i \tag{16}$$

where the index $i$ corresponds to the $i$th sphere of *M* spheres. The number of zeroth order image charges is equal to the number of spheres, *M*.

The first order image charges comparable to Eqs. (4) can be expressed as:

$$q_{ij}_{j \neq i} = -\frac{a_i}{|\mathbf{r}_i - \mathbf{r}_j|} q_j \tag{17a}$$

and is located at:

$$\mathbf{r}_{ij}_{j \neq i} = \mathbf{r}_i - a_i^2 \frac{\mathbf{r}_i - \mathbf{r}_j}{|\mathbf{r}_i - \mathbf{r}_j|^2} \tag{17b}$$

where $\mathbf{r}_i$ is the location of the $i$th sphere. The interpretation of Eq. (17a) is that $q_{ij}$ is the first order image charge inside of sphere $i$ located at $\mathbf{r}_{ij}$ that is induced by the zeroth order image charge inside of sphere $j$. The number of first order image charges is equal to $M(M-1)$, since for each of the *M* spheres, there are *M*

- 1 first order images contained in that sphere. Note that in the special case where the *j*th sphere is grounded (i.e., the *j*th sphere has no zeroth order charge), the first order image charge induced in the *i*th sphere by the *j*th sphere is also zero.

Now consider the second order image charges, and this time there are $M(M-1)^2$ possibilities:

$$q_{ijk \atop {j \neq i \atop k \neq j}} = -\frac{a_i}{|\mathbf{r}_i - \mathbf{r}_{jk}|} q_{jk} \tag{18a}$$

located at:

$$\mathbf{r}_{ijk \atop {j \neq i \atop k \neq j}} = \mathbf{r}_i - a_i^2 \frac{\mathbf{r}_i - \mathbf{r}_{jk}}{|\mathbf{r}_i - \mathbf{r}_{jk}|^2} \tag{18b}$$

As in the first order case, the interpretation is that $q_{ijk}$ is the second order image charge inside of sphere *i*, located at $\mathbf{r}_{ijk}$, that is induced by the first order image charge $q_{jk}$ located at $\mathbf{r}_{jk}$. The total number of images charges *L* including *N*th order for *M* spheres is:

$$L \leq M \sum_{n=0}^{N} (M-1)^n \tag{19}$$

The inequality considers the case where one or more spheres may be grounded, in which case the number of images charges is reduced, as previously discussed.

Using the above definitions for image charges and their positions the electric potential for *M* conducting spheres, analogous to Eq. (7), can be expressed by the following series:

$$U(\mathbf{r}) = \frac{1}{4\pi\varepsilon_0} \sum_{i=1}^{M} \left( \frac{q_i}{|\mathbf{r} - \mathbf{r}_i|} + \sum_{j=1 \atop j \neq i}^{M} \left( \frac{q_{ij}}{|\mathbf{r} - \mathbf{r}_{ij}|} + \sum_{k=1 \atop k \neq j}^{M} \left( \frac{q_{ijk}}{|\mathbf{r} - \mathbf{r}_{ijk}|} + \cdots \right) \right) \right) \tag{20}$$

In the two sphere case, Eqs. (20) and (7) are equivalent, as they should be, for the same image order. This can be easily seen by noting that the summations in Eq. (20) reduce to single terms for *M* = 2.

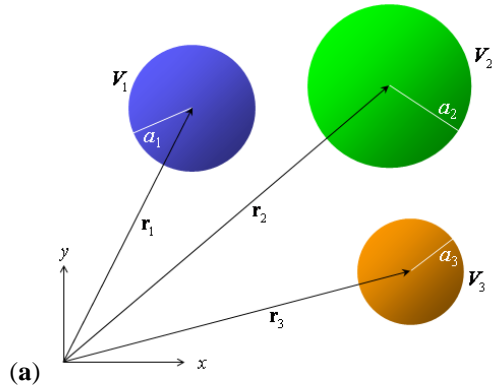

(a)

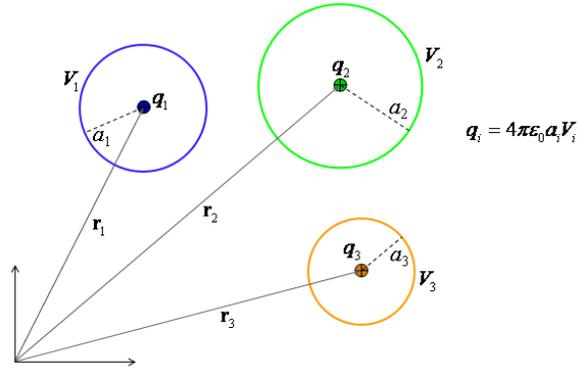

All Zeroth Order Image Charges

(b) $M$ = Number of Spheres = 3    Number of $0^{th}$ Order Charges = $M$ = 3

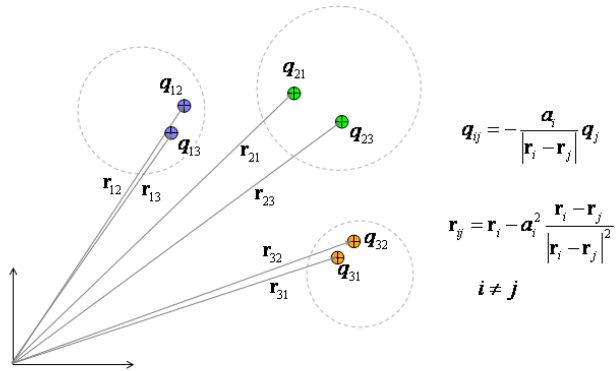

All First Order Image Charges

(c) $M$ = Number of Spheres = 3    Number of $1^{st}$ Order Charges = $M(M-1)$ = 6

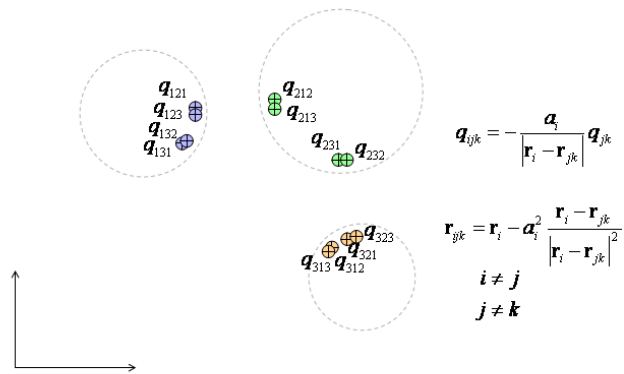

All Second Order Image Charges

(d) $M$ = Number of Spheres = 3    Number of $2^{nd}$ Order Charges = $M(M-1)^2$ = 12

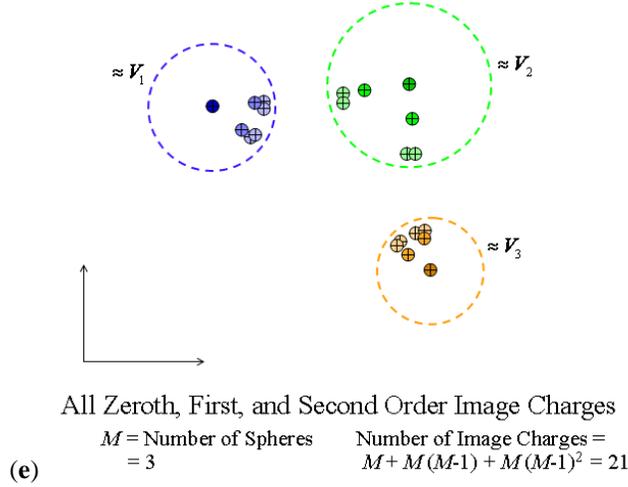

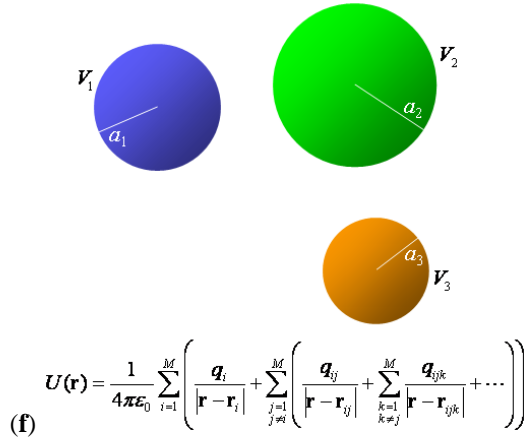

(f) $$U(\mathbf{r}) = \frac{1}{4\pi\varepsilon_0} \sum_{i=1}^{M} \left( \frac{q_i}{|\mathbf{r} - \mathbf{r}_i|} + \sum_{\substack{j=1 \\ j \neq i}}^{M} \left( \frac{q_{ij}}{|\mathbf{r} - \mathbf{r}_{ij}|} + \sum_{\substack{k=1 \\ k \neq j}}^{M} \frac{q_{ijk}}{|\mathbf{r} - \mathbf{r}_{ijk}|} + \cdots \right) \right)$$

**Fig. (7).** Demonstration of image charge determinations for $M$ conducting spheres.: (**a**) $M = 3$ example; (**b**) zeroth order images using Eq. (16); (**c**) first order images using Eqs. (17); (**d**) second order images using Eqs. (18); (**e**) all zeroth, first, and second order image charges; (**f**) truncated potential from Eq. (20) for $M = 3$ spheres and order $N = 2$.

**Charge Normalization of M Conducting Spheres**

Optimization of the potential in Eq. (20) by charge normalization proceeds in a manner similar to that previously discussed for the two sphere case. Analogous to Eqs. (9), the voltages of the zeroth order charges are adjusted by scaling factors $\alpha_i$:

$$q_i = 4\pi\varepsilon_0 a_i (\alpha_i V_i) \tag{21}$$

A truncated solution given by Equation (20) can be optimized by adjusting the $\alpha_i$'s so that the voltage on the sphere surface $U(\mathbf{r} = \mathbf{r}_i + a_i \mathbf{e}) \to V_i$ for $i = 1$ to $M$. This can be summarized by the following matrix equation:

$$\begin{pmatrix} U(\mathbf{r}_1 + a_1\mathbf{e}) \\ U(\mathbf{r}_2 + a_2\mathbf{e}) \\ \vdots \\ U(\mathbf{r}_M + a_M\mathbf{e}) \end{pmatrix} = \begin{pmatrix} V_1 \\ V_2 \\ \vdots \\ V_M \end{pmatrix} = \mathbf{C} \cdot \begin{pmatrix} \alpha_1 \\ \alpha_2 \\ \vdots \\ \alpha_M \end{pmatrix} \qquad (22)$$

where the components of **C** are obtained from Equation (20) by re-ordering terms so that the $q_i$'s can be grouped and factored. The reordering is accomplished by first reversing the order of the indices of the $q$'s in Equation (20), substituting in Eqs. (17a) and (18a), then factoring out $q_i$. For the second order case, this proceeds as follows:

$$\begin{aligned} U(\mathbf{r}_l + a_l\mathbf{e}) &= \frac{1}{4\pi\varepsilon_0} \sum_{i=1}^{M} \left( \frac{q_i}{|\mathbf{r}_l + a_l\mathbf{e} - \mathbf{r}_i|} + \sum_{\substack{j=1 \\ j \neq i}}^{M} \left( \frac{q_{ij}}{|\mathbf{r}_l + a_l\mathbf{e} - \mathbf{r}_{ij}|} + \sum_{\substack{k=1 \\ k \neq j}}^{M} \frac{q_{ijk}}{|\mathbf{r}_l + a_l\mathbf{e} - \mathbf{r}_{ijk}|} \right) \right) \\ &= \frac{1}{4\pi\varepsilon_0} \sum_{i=1}^{N} \left( \frac{q_i}{|\mathbf{r}_l + a_l\mathbf{e} - \mathbf{r}_i|} + \sum_{\substack{j=1 \\ j \neq i}}^{N} \left( \frac{q_{ji}}{|\mathbf{r}_l + a_l\mathbf{e} - \mathbf{r}_{ji}|} + \sum_{\substack{k=1 \\ k \neq j}}^{N} \frac{q_{kji}}{|\mathbf{r}_l + a_l\mathbf{e}_l - \mathbf{r}_{kji}|} \right) \right) \\ &= \frac{1}{4\pi\varepsilon_0} \sum_{i=1}^{N} \left( \frac{q_i}{|\mathbf{r}_l + a_l\mathbf{e} - \mathbf{r}_i|} + \sum_{\substack{j=1 \\ j \neq i}}^{N} q_{ji} \left( \frac{1}{|\mathbf{r}_l + a_l\mathbf{e} - \mathbf{r}_{ji}|} + \sum_{\substack{k=1 \\ k \neq j}}^{N} \frac{-a_k}{|\mathbf{r}_k - \mathbf{r}_{ji}| \cdot |\mathbf{r}_l + a_l\mathbf{e} - \mathbf{r}_{kji}|} \right) \right) \\ &= \frac{1}{4\pi\varepsilon_0} \sum_{i=1}^{N} q_i \left( \frac{1}{|\mathbf{r}_l + a_l\mathbf{e} - \mathbf{r}_i|} + \sum_{\substack{j=1 \\ j \neq i}}^{N} \left( \frac{-a_j}{|\mathbf{r}_j - \mathbf{r}_i| \cdot |\mathbf{r}_l + a_l\mathbf{e} - \mathbf{r}_{ji}|} + \sum_{\substack{k=1 \\ k \neq j}}^{N} \frac{a_j a_k}{|\mathbf{r}_j - \mathbf{r}_i| \cdot |\mathbf{r}_k - \mathbf{r}_{ji}| \cdot |\mathbf{r}_l + a_l\mathbf{e} - \mathbf{r}_{kji}|} \right) \right) \\ &= \sum_{i=1}^{N} c_{li} \end{aligned} \qquad (23)$$

Analogous to Eq. (13), the components of **C** when averaged over the surface of sphere $l$ are:

$$C_{li} = \frac{1}{N_R} \sum_{n=1}^{N_R} c_{li}(s_n\pi, u_n 2\pi) \qquad (24)$$

where,

$$c_{li}(\theta, \phi) = a_i V_i \left( \frac{1}{|\mathbf{r}_l + a_l\mathbf{e} - \mathbf{r}_i|} + \sum_{\substack{j=1 \\ j \neq i}}^{N} \left( \frac{-a_j}{|\mathbf{r}_j - \mathbf{r}_i| \cdot |\mathbf{r}_l + a_l\mathbf{e} - \mathbf{r}_{ji}|} + \sum_{\substack{k=1 \\ k \neq j}}^{N} \frac{a_j a_k}{|\mathbf{r}_j - \mathbf{r}_i| \cdot |\mathbf{r}_k - \mathbf{r}_{ji}| \cdot |\mathbf{r}_l + a_l\mathbf{e} - \mathbf{r}_{kji}|} \right) \right) \qquad (25)$$

and $s_n$ is defined by Eq. (14). Finally, the $\alpha_i$'s are found by inverting **C**, similar to Eq. (15):

$$\begin{pmatrix} \alpha_1 \\ \alpha_2 \\ \vdots \\ \alpha_M \end{pmatrix} = \mathbf{C}^{-1} \cdot \begin{pmatrix} V_1 \\ V_2 \\ \vdots \\ V_M \end{pmatrix} \qquad (26)$$

An example is given in Figs. (**8**), using three spheres with the following parameters: $V_1 = 0.2$, $V_2 = 0.8$, $V_3 = -0.5$, $a_2 = 1.0$, $a_3 = 0.7$, $\mathbf{r}_1 = (0, 0, 0)$, $\mathbf{r}_2 = (d, 0, 0)$, $\mathbf{r}_3 = (0, d, 0)$, $a_1 = 1.5$, and $d = 3.5$. The significant improvement due to charge normalization is readily apparent.

(**a**)

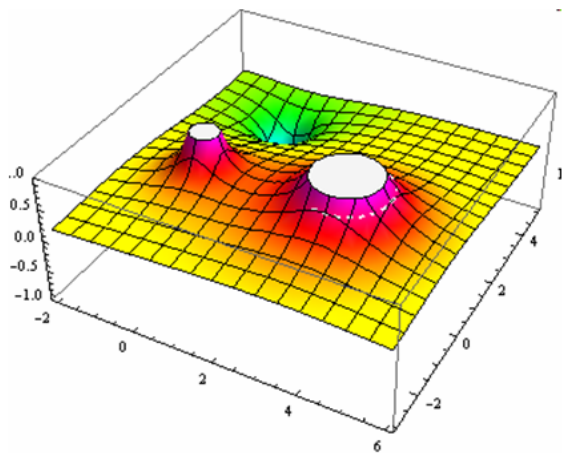 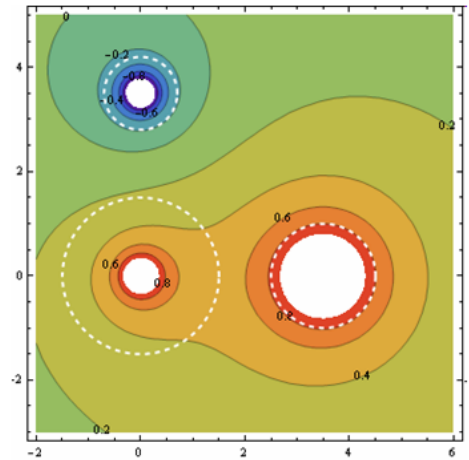

(**b**)

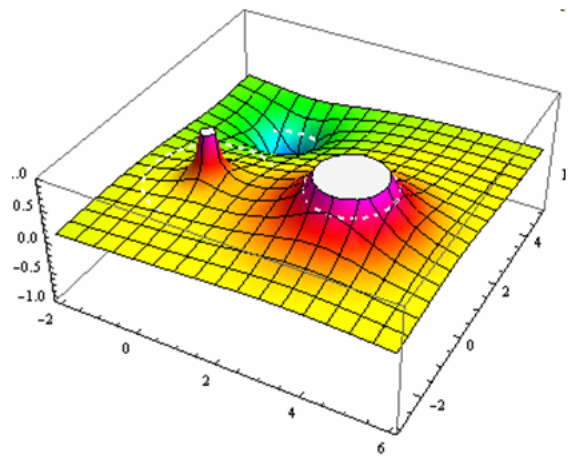 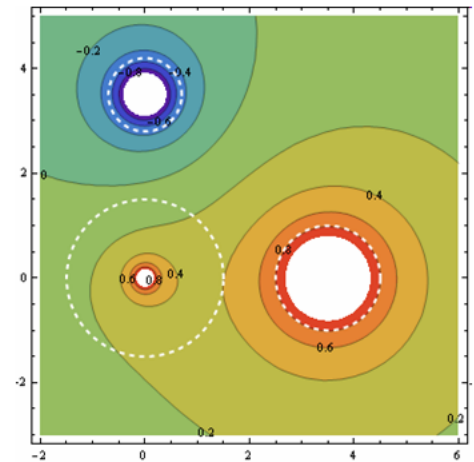

(**c**)

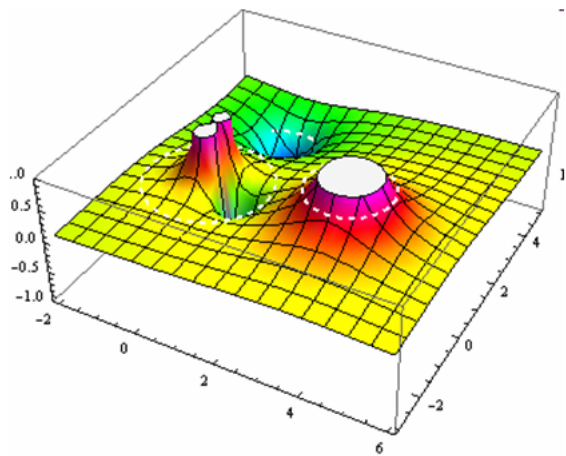 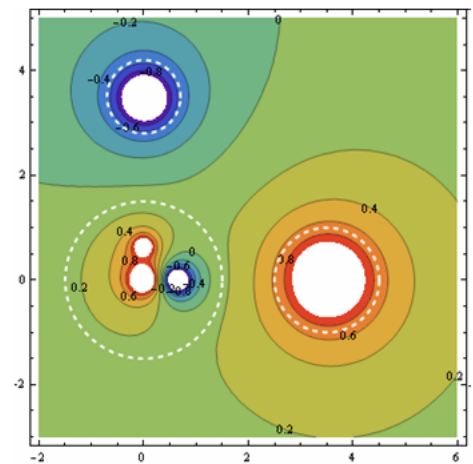

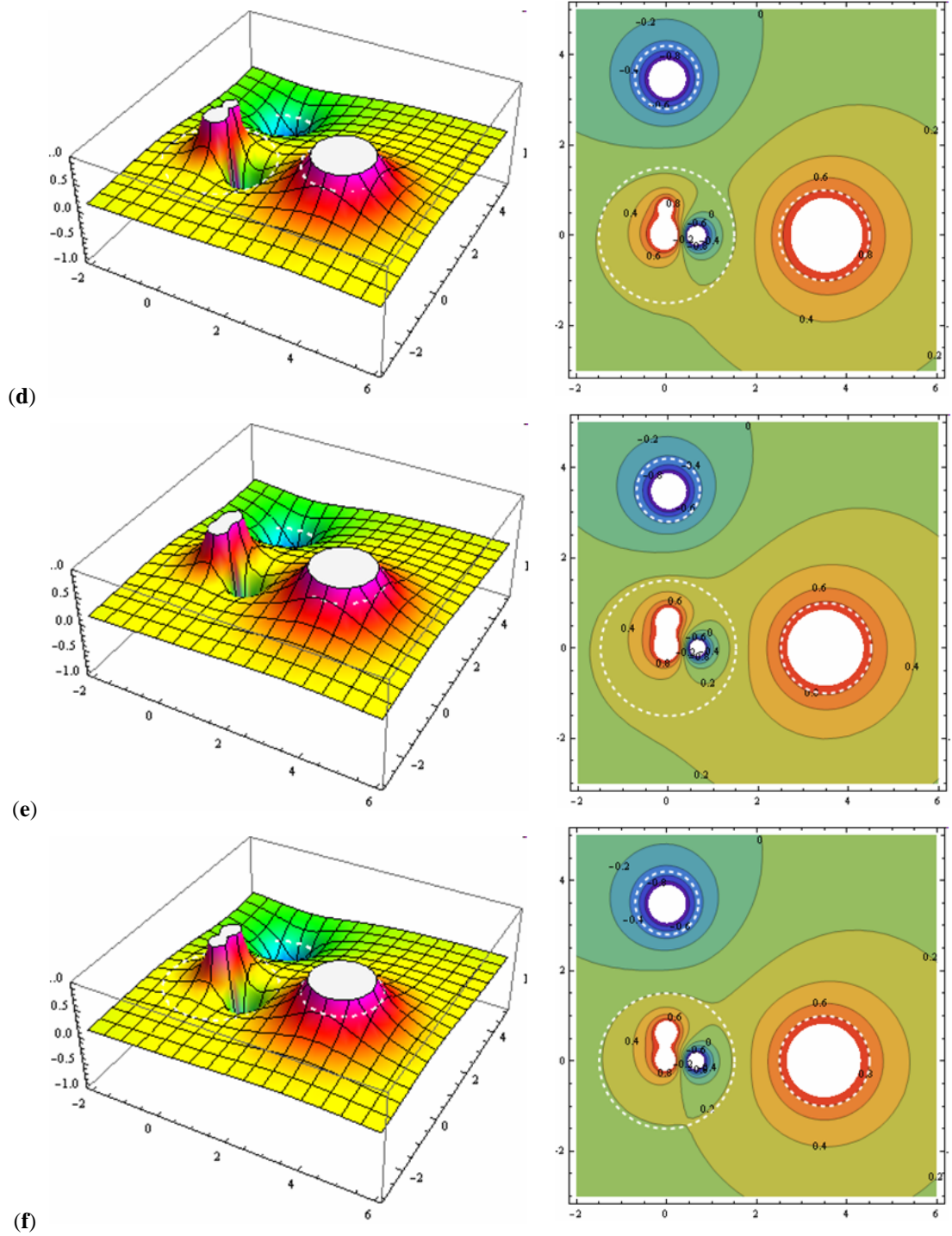

**Fig. (8).** Comparison of zeroth order through second order image method electric field potentials of three conducting spheres. The potential $U(\mathbf{r})$ is pseudo colorized, and is computed from Eq. (20) at $z = 0$ along the $x$-$y$ plane without and with charge normalization using Eqs. (24) through (26). The dotted white circle represents the correct potential on the sphere surfaces: (**a**) zeroth order image term only; (**b**) zeroth order term with normalization; (**c**) include first order image term; (**d**) including first order terms with normalization; (**e**) includes second order image order term; (**f**) including second order terms with normalization.

## GENERAL IMAGE CHARGE OPTIMIZATION

The charge normalization described by Eq. (26) is a special case of a parameter optimization procedure where only the magnitude of the zeroth order charge is modified so that the calculated electrical field potential matches the actual voltage on all sphere surfaces. A more general approach is to parameterize the position of each charge as well as its magnitude. With this approach comes the philosophy that image charges are not necessarily intrinsic to a solution, but simply provide an initial guess for position and magnitude of some number of charges inside each sphere that will approximate a solution to Laplace's equation, in all space and on the surface of the spheres. The potential field due to a system of $M$ conducting spheres can be approximated by $L$ point charges, where $L \geq M$. The potential from a system of $L$ point charges is then:

$$U(\mathbf{r}) = \sum_{j=1}^{L} \frac{\beta_j}{|\mathbf{r}-\mathbf{r}_j|}$$

$$= \sum_{j=1}^{L} \frac{\beta_j}{\sqrt{(x-x_j)^2 + (y-y_j)^2 + (z-z_j)^2}} \quad (27)$$

In the case of $L = M$, the numerator of Eq. (27) is due to the zeroth image charge of the $i$th sphere, as described by Eq. (21), so that $\beta_i = \frac{q_i}{4\pi\varepsilon_0} = a_i(\alpha_i V_i)$.

The means to a general optimization approach is to define an error function. It is imperative to select a metric that when maximized or minimized will ensure that the field more accurately represents the exact field of the non-truncated series in a relevant way. In this sense, "accurately" may be difficult to define, but we show below that it is made easier to define by noting that the electric field in all space will be correct as long as the electric field on the surface of the spheres themselves is correct. Also, (as demonstrated below) it is obvious along the surface of the charged spheres that the field resulting from a truncated series of image charges is qualitatively different from the correct field, whereas the field resulting from an optimized set of charges is qualitatively more similar to the correct field. Therefore, we should expect in almost all applications that the field resulting from the optimized set of charges will more "accurately" represent the field in the relevant way. For example, it would appear obvious seeing the qualitative improvement that the calculated trajectories of space radiation particles will follow the correct trajectories more closely when the field is qualitatively correct rather than qualitatively incorrect. We thus propose to use a *least squares* of the value of electric potential as the metric, and we calculate it only upon the surface of the charged spheres. We demonstrate below the quantitative improvement of this metric via the Gradient Search Method [13].

By defining an error $E$, based on the sum of potentials on the surface over all $M$ spheres, the *best fit* $\beta_j$ and $\mathbf{r}_j$ in Eq. (27) can be determined by minimizing the error function:

$$E = \frac{1}{4\pi} \sum_{i=1}^{M} \int_0^{2\pi} d\phi \int_0^{\pi} [V_i - U_i(\theta,\phi)]^2 \sin\theta \, d\theta \quad (28)$$

where,

$$U_i(\theta,\phi) = \sum_{j=1}^{L} \frac{\beta_k}{\left[(x_i + a_i \sin\theta\cos\phi - x_j)^2 + (y_i + a_i \sin\theta\sin\phi - y_j)^2 + (z_i + a_i \cos\theta - z_j)^2\right]^{1/2}} \quad (29)$$

The integral over the spherical angles $\theta$ and $\phi$ in Eq. (28) is more formal way of describing the averaging over $N_R$ points described in Eq. (13) and again by Equation (24). The *Monte Carlo* method of randomly picking points on the surface for evaluation can also be used here to perform the integration.

Equation (28) can be minimized by searching for the zero of its gradient:

$$\nabla E = \begin{pmatrix} \frac{\partial}{\partial \beta_1} \\ \vdots \\ \frac{\partial}{\partial x_1} \\ \vdots \\ \frac{\partial}{\partial y_1} \\ \vdots \\ \frac{\partial}{\partial z_1} \\ \vdots \end{pmatrix} E(\mathbf{P}) \to 0 \quad , \text{where} \quad \mathbf{P} = \begin{pmatrix} \beta_1 \\ \vdots \\ \beta_L \\ x_1 \\ \vdots \\ x_L \\ y_1 \\ \vdots \\ y_L \\ z_1 \\ \vdots \\ z_L \end{pmatrix} \tag{30}$$

The error $E(\mathbf{P})$ is a function of $4L$ parameters, designated by the parameter vector $\mathbf{P}$. The problem is to find the minimum of $E(\mathbf{P})$ in $4L$ dimensional parameter space. (The difficulties associated with finding a global versus local minimum are intrinsic to gradient search methods and will not be discussed here.) The gradient search is implemented by the following recursion relation:

$$\mathbf{P}(n+1) = \mathbf{P}(n) - \mu \nabla E(n) \tag{31}$$

where $n$ denotes the $n$th step of the recursion, and $\mu$ is the *convergence constant*. Note that $\mu$ can be determined empirically and is not necessarily a constant. For all time steps $n$, if $\mu$ is too small, convergence will require an unreasonably large number of iterations of Eq. (31). If $\mu$ is too large, Eq. (31) may diverge, usually towards infinity, or oscillate wildly about the solution.

The gradient of the error function in Eq. (31) is computed as follows:

$$\begin{aligned} \nabla E &= \nabla \left\{ \frac{1}{4\pi} \sum_{i=1}^{M} \int_0^{2\pi} d\phi \int_0^{\pi} [V_i - U_i(\theta,\phi)]^2 \sin\theta \, d\theta \right\} \\ &= \frac{1}{4\pi} \sum_{i=1}^{M} \int_0^{2\pi} d\phi \int_0^{\pi} \nabla \{ [V_i - U_i(\theta,\phi)]^2 \} \sin\theta \, d\theta \\ &= -\frac{1}{2\pi} \sum_{i=1}^{M} \int_0^{2\pi} d\phi \int_0^{\pi} [V_i - U_i(\theta,\phi)] \nabla \{ U_i(\theta,\phi) \} \sin\theta \, d\theta \end{aligned} \tag{32}$$

The gradient term in Equation (32) is a vector:

$$\nabla\{U_i(\theta,\phi)\} \equiv \mathbf{a}_i = \begin{pmatrix} (\alpha_\beta)_1 \\ \vdots \\ (\alpha_\beta)_L \\ (\alpha_x)_1 \\ \vdots \\ (\alpha_x)_L \\ (\alpha_y)_1 \\ \vdots \\ (\alpha_y)_L \\ (\alpha_z)_1 \\ \vdots \\ (\alpha_z)_L \end{pmatrix}_i \tag{33}$$

where,

$$\left((\alpha_\beta)_j\right)_i = \frac{\partial U_i(\theta,\phi)}{\partial \beta_j}$$
$$= \frac{1}{\left[(x_i + a_i \sin\theta\cos\phi - x_j)^2 + (y_i + a_i \sin\theta\sin\phi - y_j)^2 + (z_i + a_i \cos\theta - z_j)^2\right]^{1/2}} \tag{34a}$$

$$\left((\alpha_x)_j\right)_i = \frac{\partial U_i(\theta,\phi)}{\partial x_j}$$
$$= \frac{2\beta_j(x_i + a_i \sin\theta\cos\phi - x_j)}{\left[(x_i + a_i \sin\theta\cos\phi - x_j)^2 + (y_i + a_i \sin\theta\sin\phi - y_j)^2 + (z_i + a_i \cos\theta - z_j)^2\right]^{1/2}} \tag{34b}$$

$$\left((\alpha_y)_j\right)_i = \frac{\partial U_i(\theta,\phi)}{\partial y_j}$$
$$= \frac{2\beta_j(y_i + a_i \sin\theta\cos\phi - y_j)}{\left[(x_i + a_i \sin\theta\cos\phi - x_j)^2 + (y_i + a_i \sin\theta\sin\phi - y_j)^2 + (z_i + a_i \cos\theta - z_j)^2\right]^{1/2}} \tag{34c}$$

$$\left((\alpha_z)_j\right)_i = \frac{\partial U_i(\theta,\phi)}{\partial z_j}$$
$$= \frac{2\beta_j(z_i + a_i \sin\theta\cos\phi - z_j)}{\left[(x_i + a_i \sin\theta\cos\phi - x_j)^2 + (y_i + a_i \sin\theta\sin\phi - y_j)^2 + (z_i + a_i \cos\theta - z_j)^2\right]^{1/2}} \tag{34d}$$

Figs. (**9**) revisits the three sphere example, including the zeroth, first, and second order normalized image charge solutions, using the same parameters from the Figs. (**8**). For comparison, the three zeroth order charges are optimized by the gradient search algorithm, with $L = M = 3$. A second gradient search case is included for comparison with $L = 4$, where two charges are located within sphere-1. By implementing the gradient search algorithm described by Equation (31), the point charges change in magnitude and position to minimize the error function defined by Eq. (28). The final magnitude and charge positions are given by the $\mathbf{P}(n)$ vector from Equation (30). The initial $\mathbf{P}(0)$ vector is determined by a set of image charges, three zeroth order and one first order charge. The final solution vector $\mathbf{P}(n)$ is somewhat arbitrarily determined to be solution when the error falls below some predefined value, are the change in error between iterations

converges to zero. For comparison, the gradient search case in Fig. (**9d**) with *L* = 4 is qualitatively better than the full first order case of Fig. (**9c**) where the total number of image charges is equal to 9, and not as good as the full second order case of Fig. (**9e**) where the total number of charges is equal to 21.

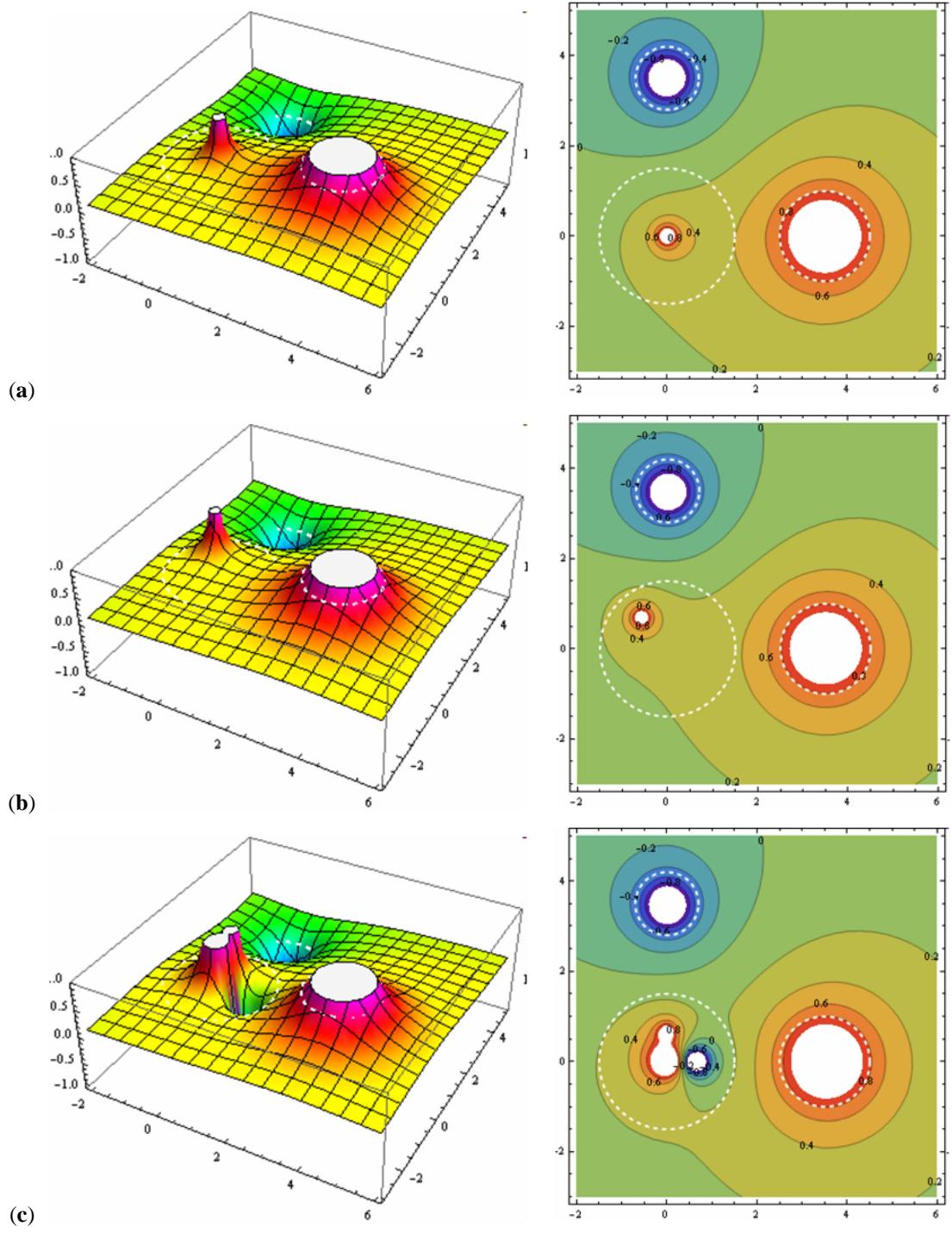

(**a**)

(**b**)

(**c**)

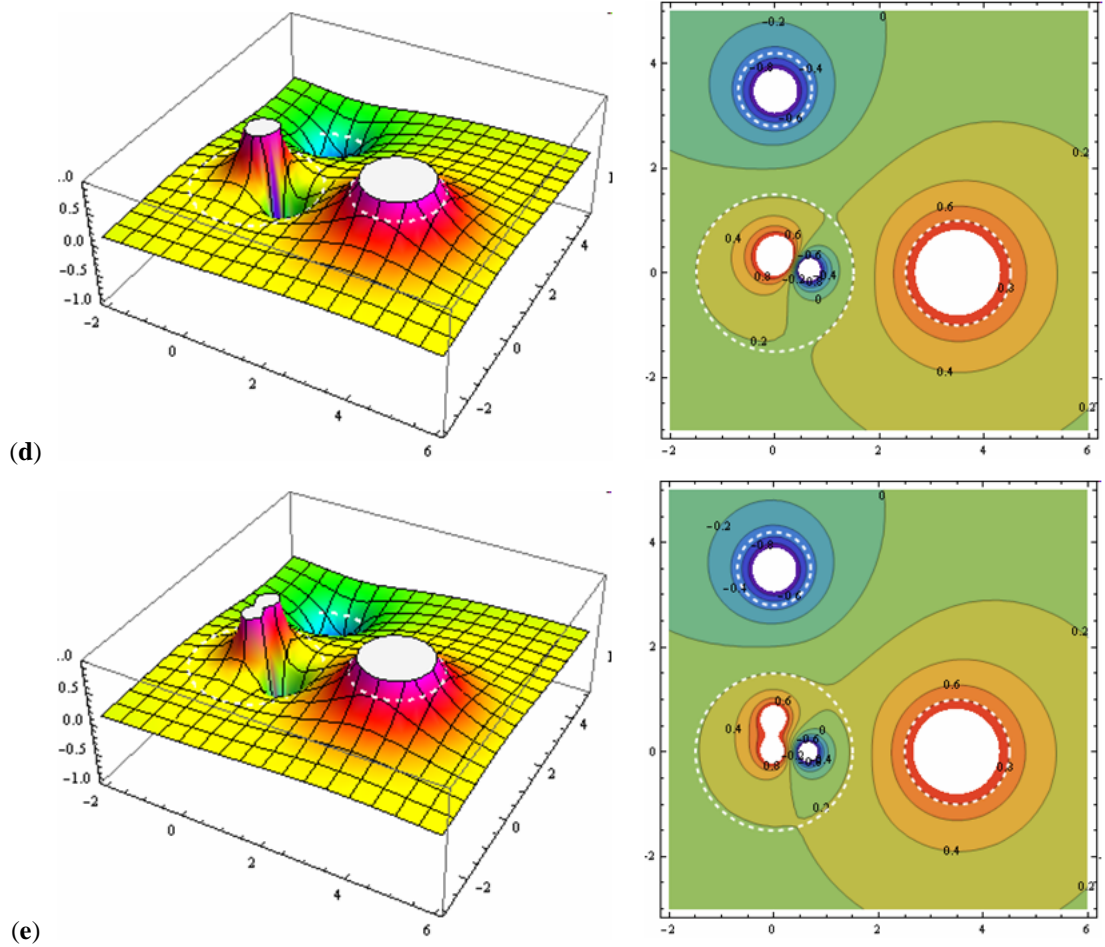

**Fig. (9).** Zeroth through second order electric potential solutions from three conducting spheres using normalized image charges compared to a general optimization with a gradient search algorithm. The potential $U(\mathbf{r})$ is pseudo colorized, and is computed from Eq. (20) at $z = 0$ along the $x$-$y$ plane without with charge normalization using Eqs. (21) and (26). The dotted white circle represents the correct potential on the sphere surfaces: (**a**) zeroth order normalized image charges ($L = 3$); (**b**) zeroth order normalized image charges with gradient search optimization, $L = 3$; (**c**) include first order normalized images ($L = 9$); (**d**) initialized gradient search with $L = 4$; (**e**) includes second order image order term with normalization ($L = 21$).

## SUMMARY AND DISCUSSION

It is desirable to improve the accuracy of the estimated electrical field and its energy when using the image charge method. The non-truncated series of image charges gives the exact solution, so we wish to improve the accuracy of the truncated series of image charges so that it will better approximate the fields that would have resulted from a non-truncated series. Accuracy can be improved by three methods:

(1) Increase the length of the series of image charges (i.e., include increasingly higher order charges). This is the straight-forward method, of course. It works because the series is an alternating series and converges to the correct electric field as the order $N \to \infty$, and thus the magnitude of the error in the energy of the field decreases to zero.

(2) Truncating the series results in a set of image charges for each sphere that do not sum to the correct overall charge of the sphere. A simple improvement is therefore to renormalize the total charge by rescaling all the image charges associated with one sphere by the same factor so that they do sum to the correct overall charge of the sphere. In the case of spheres held at a constant potential, the normalization is implemented to match the specified potential. This procedure is a special case of optimization and produces significant improvement for low order truncations.

(3) Optimize a system of $L$ image charges for $M$ spheres by (i) moving the point charges individually away from the locations that were specified by the image charge series and (ii) redistributing the charge between them so that the amount of charge for each point is different than was initially specified by the image charge series. This may be accomplished by a gradient search algorithm that seeks to optimize some metric of the fidelity of the overall field. This method is possible because a set of $N$ point charges in a truncated series is not necessarily more accurate at approximating the non-truncated series as is some other set of $L$ point charges where $L < M$. This may seem counter-intuitive at first, but upon reflection there is no reason why the series truncated to $N$ point charges should be the most accurate solution. The only feature that makes the first $N$ terms in the image charge series special is that they contribute to the exact solution when an infinite set of additional points in the series is included. However, if we know a priori that only a finite number of point charges are going to be included, then there is no reason to expect the first $N$ terms of a truncated series to be the best solution. This concept is generally true in all practical applications of infinite mathematical series. For example, transcendental functions implemented on a computer often make use of this characteristic of truncated infinite series.

**REFERENCES**

[1] Maxwell JC. A treatise on electricity and magnetism. 3$^{rd}$ ed. Oxford: Clarendon Press; 1891.
[2] Jackson JD. Classical electrodynamics. 3$^{rd}$ ed. New York: John Wiley and Sons; 1998.
[3] Reitz JR, Milford FJ. Foundations of electromagnetic theory. 2$^{nd}$ ed. Addison-Wesley Publishing Company; 1967.
[4] Hayt WH. Engineering electromagnetics. 3$^{rd}$ ed. New York: McGraw-Hill Book Company; 1974.
[5] Schinzinger R, Laura PAA. Conformal mapping: methods and applications. Courier Dover Publications; 2003.
[6] Tikhonov, AN, Samarskii, AA. Equations of mathematical physics. reprint. Courier Dover Publications; 1990.
[7] Crowdy, DG. Schwarz–Christoffel mappings to unbounded multiply connected polygonal regions. Math Camb Phil Soc 2007; 142: 319–21.
[8] DeLillo, TK. Schwarz–Christoffel mapping of bounded, multiply connected domains. Computational Methods and Function Theory 2006; 6(2): 275–26.
[9] DeLillo, TK, Driscoll, TA, Elcrat, AR, Pfaltzgraff, JA. Computation of multiply connected Schwarz–Christoffel maps for exterior domains. Computational Methods and Function Theory 2006; 6(2): 301–15.
[10] Gibson, WC. The method of moments in electromagnetics. CRC Press; 2007.
[11] Bonderson, A, Rylander, T, Ingelström, P. Computational electromagnetics. Springer; 2005.
[12] Soules, J. Precise calculation of the electrostatic force between charged spheres including induction effects. Am J Phys 1990; 58(12): 1195-5.
[13] Press WH, Teukolsky SA, Vetterling WT, Flannery BP. Numerical recipes – the art of scientific computing. 3$^{rd}$ ed. Cambridge University Press; 2007.